\def \SAIT #1 #2 {{\em Mem.\ Soc.\ Astron.\ It.\/} {\bf #1}, #2}
\def \MESS #1 #2 {{\em The Messenger\/} {\bf #1}, #2}
\def \ASTRNACH #1 #2 {{\em Astron. Nach.\/} {\bf #1}, #2}
\def \AAP #1 #2 {{\em Astron. Astrophys.\/} {\bf #1}, #2}
\def \AAL #1 #2 {{\em Astron. Astrophys. Lett.\/} {\bf #1}, L#2}
\def \AAR #1 #2 {{\em Astron. Astrophys. Rev.\/} {\bf #1}, #2}
\def \AAS #1 #2 {{\em Astron. Astrophys. Suppl. Ser.\/} {\bf #1}, #2}
\def \AJ #1 #2 {{\em Astron. J.\/} {\bf #1}, #2}
\def \ANNREV #1 #2 {{\em Ann. Rev. Astron. Astrophys.\/} {\bf #1}, #2}
\def \APJ #1 #2 {{\em Astrophys. J.\/} {\bf #1}, #2}
\def \APJL #1 #2 {{\em Astrophys. J. Lett.\/} {\bf #1}, L#2}
\def \APJS #1 #2 {{\em Astrophys. J. Suppl.\/} {\bf #1}, #2}
\def \APSS #1 #2 {{\em Astrophys. Space Sci.\/} {\bf #1}, #2}
\def \ASR #1 #2 {{\em Adv. Space Res.\/} {\bf #1}, #2}
\def \BAIC #1 #2 {{\em Bull. Astron. Inst. Czechosl.\/} {\bf #1}, #2}
\def \JSQRT #1 #2 {{\em J. Quant. Spectrosc. Radiat. Transfer\/} {\bf #1}, #2}
\def \MN #1 #2 {{\em Mon. Not. R. Astr. Soc.\/} {\bf #1}, #2}
\def \MEM #1 #2 {{\em Mem. R. Astr. Soc.\/} {\bf #1}, #2}
\def \PLR #1 #2 {{\em Phys. Lett. Rev.\/} {\bf #1}, #2}
\def \PASJ #1 #2 {{\em Publ. Astron. Soc. Japan\/} {\bf #1}, #2}
\def \PASP #1 #2 {{\em Publ. Astr. Soc. Pacific\/} {\bf #1}, #2}
\def \NAT #1 #2 {{\em Nature\/} {\bf #1}, #2}

\documentstyle{memsait}
\input epsf.sty
\begin{opening}
\title{AUTOMATED SEARCH FOR LSB GALAXIES ON DPOSS}
\author{S. Sabatini$^1$, R. Scaramella$^1$, V. Testa$^1$, 
S. Andreon$^2$, G. Longo$^2$, G. Djorgovski$^3$, R.R. De Carvalho$^4$}
\institute{$^1$Osservatorio Astronomico di Roma, Roma, Italia\\
$^2$Osservatorio Astronomico di Capodimonte, Napoli, Italia\\
$^3$Caltech, Pasadena\\
$^4$Observatorio Nacional Rio de Janeiro, Rio de Janeiro, Brasil\\}
\date{} % DO NOT INSERT ANY DATE HERE !!!
\end{opening}

\begin{document}

%\oddpagefooter{\sf Mem. S.A.It., Vol. ??, ??}{}{\thepage}
%\evenpagefooter{\thepage}{}{\sf Mem. S.A.It., Vol. ??, ??}
\oddpagefooter{}{}{} % LEAVE AS IT IS !
\evenpagefooter{}{}{} % LEAVE AS IT IS !
\ 
\bigskip

\begin{abstract}
{\it Low Surface Brightness Galaxies} (LSBGs) today constitute the less known 
fraction of galaxies: their number density and their physical properties 
(e.g. luminosity, colour, dynamics) are still largely uncertain. This is 
mainly due to the difficoulty in finding them (since their presence goes 
often unnoticed when astronomical images are analyzed with standard methods) 
and in measuring their properties.  
%This is mainly due to the difficulty in finding them, since their properties
% are such that their presence goes often unnoticed when astronomical images 
%are analyzed with standard methods.
We present here the first results of a novel application to DPOSS plates,
 a project within the CRoNaRio collaboration. The aim is to build a large
catalog of LSBGs, with much better understanding of completness and biases. 
First results are quite encouraging: in the same regions covered by 
existing catalogues, done on the same plates, we find several new LSBGs in 
addition to the known ones.
\end{abstract}

\section{Introduction}
During the past twenty years, there has been a growing appreciation of 
the strong biases against finding galaxies of low surface brightness 
(Disney,  1976). These
 biases arise because the night sky is not particularly dark: because of 
the brightness of the sky background, the ability to detect a galaxy 
depends not only upon the integrated luminosity of the galaxy, but also 
upon the contrast with which the galaxy stands out above the Poisson 
fluctuations in the background. Thus our knowledge about the real 
galaxies population is incomplete: outside the Galaxy, the vast majority
of informations about stellar population, kinematics, dark matter content,
star formation and large scale clustering, have up to now been mostly
 obtained from high surface brightness galaxies (HSBGs) studies. 
The discovery and search of LSBGs constitutes, therefore, a fundamental 
enterprise in order to reach a better and more complete understanding 
of all these cosmological arguments.\\  
LSBGs have indeed a number of 
remarkable properties which distinguish them from the more familiar 
Hubble sequence of spirals (Impey $\&$ Bothun, 1997); 
the main ones are (1) LSBGs seem to constitute at 
least 50$\%$ of the total galaxy population in number and this has strong 
implications on the faint end slope of the luminosity function, on the 
baryonic matter density and expecially on galaxy formation scenarios; (2)
LSBGs discs are among the less evolved objects in the local universe since 
they have very slow SFR; (3) LSBGs are embedded
in dark matter halos which are of lower density and more extended than 
the haloes around HSBGs and are strongly dominated by dark matter 
at all radii.
%showing a systematic increase in M/L with central surface brightness.         

\section{Method}
To search for galaxies with surface density flux close to the sky noise 
level requires first an image manipulation. Detection of LSBGs using 
standard algorithms which select connected pixels above a threshold
 on the original images fails, due to the low signal to noise ratio of
these objects. Preliminar plate processing is indeed needed to obtain a 
final image where S/N ratio is enhanced.
  
The algorithm we developed for detecion of LSBGs, consists of different steps 
which can be summarized as follows:\\
%\begin{itemize}
%\item 
\indent (1) filtering of large scale background fluctuations\\   
%\item 
\indent (2) removal of stars and standard astronomical objects\\
%\item 
\indent (3) convolution of the cleaned image with {\it ad hoc} filters\\ 
%\item 
\indent (4) classification of the candidate objects detected in the previuos step.\\
%\end{itemize}
A very important consideration for LSBGs detection is to ensure
that the sky is as smooth as possible across each region of study. 
In order to achieve the required flatness we use a procedure which is part of
the package SExtractor: the technique consists in 
creating a map of the background sky in which each pixel takes 
the mean value (the mode for crowded fields) of the pixels in the original 
frame in an $n\times n$ box 
surrounding the pixel in question. This map is then subtracted from the 
original data, producing a sky-subtracted image. It should be noticed 
that any object whose size is equal to or grater than the filter box 
size will be lost, or at least severely degraded. We chose a mesh size of 
$128\times128$ pixels on DPOSS so to preserve objects having dimensions up to 2 
{\it arcmins}. 

Removal of bright stars and standard astronomical objects is a very 
important step to be done before performing convolutions: the image must be
cleaned from objects that could simulate LSBGs when convolved with the 
filters.\\
Removal is performed by replacing these objects with the local mean sky 
value taken with its noise. We use SExtractor to detect all 
the standard objects in the plate and select the ones to reject using 
a criterion based on dimension (isophotal area) and peak flux (surface 
isophotal flux weighted by peak flux), that clearly discriminates  
a stellar locus, a region of saturation and a region occupied by
diffuse objects like galaxies. We then clean the objects and performe
several convolutions of the cleaned image with 
purposely designed filters. We apply a combination of filters over different 
scales (as LSBGs have a wide range of possible scale lenght) and obtain a 
final significance image where every different scale is emphasized at
the same time. We use this image as a map of positions of candidate LSBGs.
The filters we implemented have a compensated exponential profile 

After this procedure, in the significance map LSBGs' have very high 
signal-to-noise ratio. However the sample of candidates obtained in this 
way is still contaminated. Subsequent discrimination between good 
candidates and spurious objects and classification is done by 
studying radial luminosity profiles, since LSBGs discs have mainly exponential 
profiles. 

Computer simulations were used to determine the efficiency of the method 
and were performed in two different ways: first by using already known LSBGs
 and submerging them in a progressively growing sky noise; second by 
adding on the original images artificial LSBGs characterized by defined central 
surface brightness ($\mu_\circ$) and scale lenght ($\alpha$). At present time, 
the classification efficiency is lower than the detection one, as  
the classification is performed on the original more noisier images and 
work is in progress to improve this aspect.  

\section{First results and advantages of the method}
Data used in this work consist of photographic fields (75 square 
degrees analyzed until now) from DPOSS plates in three filters 
(photographic J,F,N).
\begin{figure}
\epsfysize=9cm % fix the y-dimension and scales x-dim. to y-dim.
%\epsfxsize=8cm % fix the x-dimension and scales y-dim. to x-dim.
% Feel free to do the choice you prefer but do not exceed the x-dimension
% of the text lines
\hspace{0cm}\epsfbox{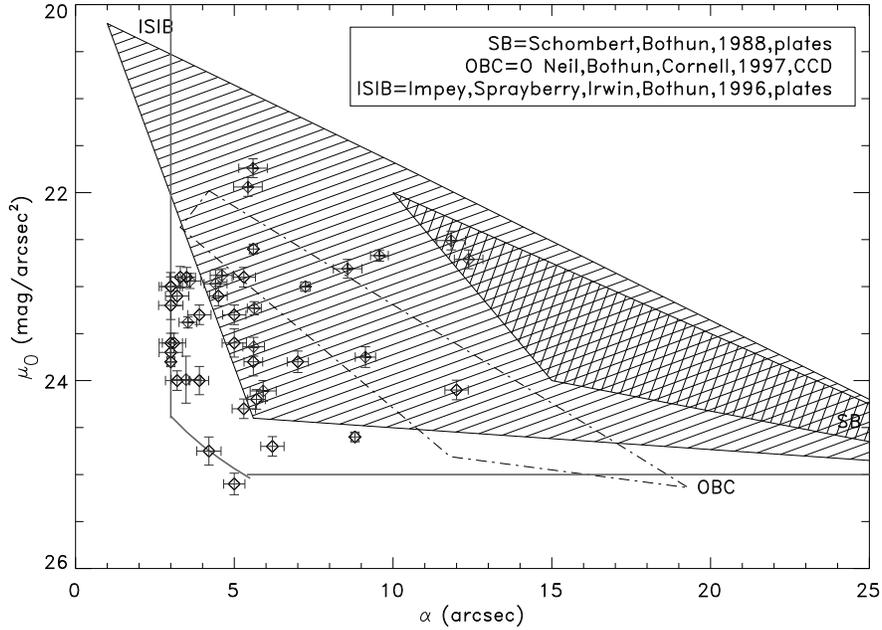} %for centering: act on hspace argument 
\caption[h]{LSBGs' luminosity profiles are mainly well fitted by an 
exponential law, with two parameters: central surface brightness
($\mu_\circ$) and scale lenght ($\alpha$). We report the regions
 covered by previous surveys as a function of these two parameters (these 
are the most important existing catalogues of LSBG), the detection limit of
our method (solid line) and the objects we have already found on the three 
DPOSS plates firstly analysed.}{\label{fig2}}
\end{figure}
In fig. \ref{fig2} we report the regions covered by previous surveys (the 
three most important existing catalogues of LSBGs) as 
a function of the two structural parameters ($\alpha$,$\mu_\circ$)
compared with our results. 
\begin{figure}
\epsfysize=6cm % fix the y-dimension and scales x-dim. to y-dim.
%\epsfxsize=8cm % fix the x-dimension and scales y-dim. to x-dim.
% Feel free to do the choice you prefer but do not exceed the x-dimension
% of the text lines
%%%\hspace{7.5cm}\epsfbox{figure/j549_17_03lsb4-prova.ps} 
\hspace{9cm}
\epsfbox{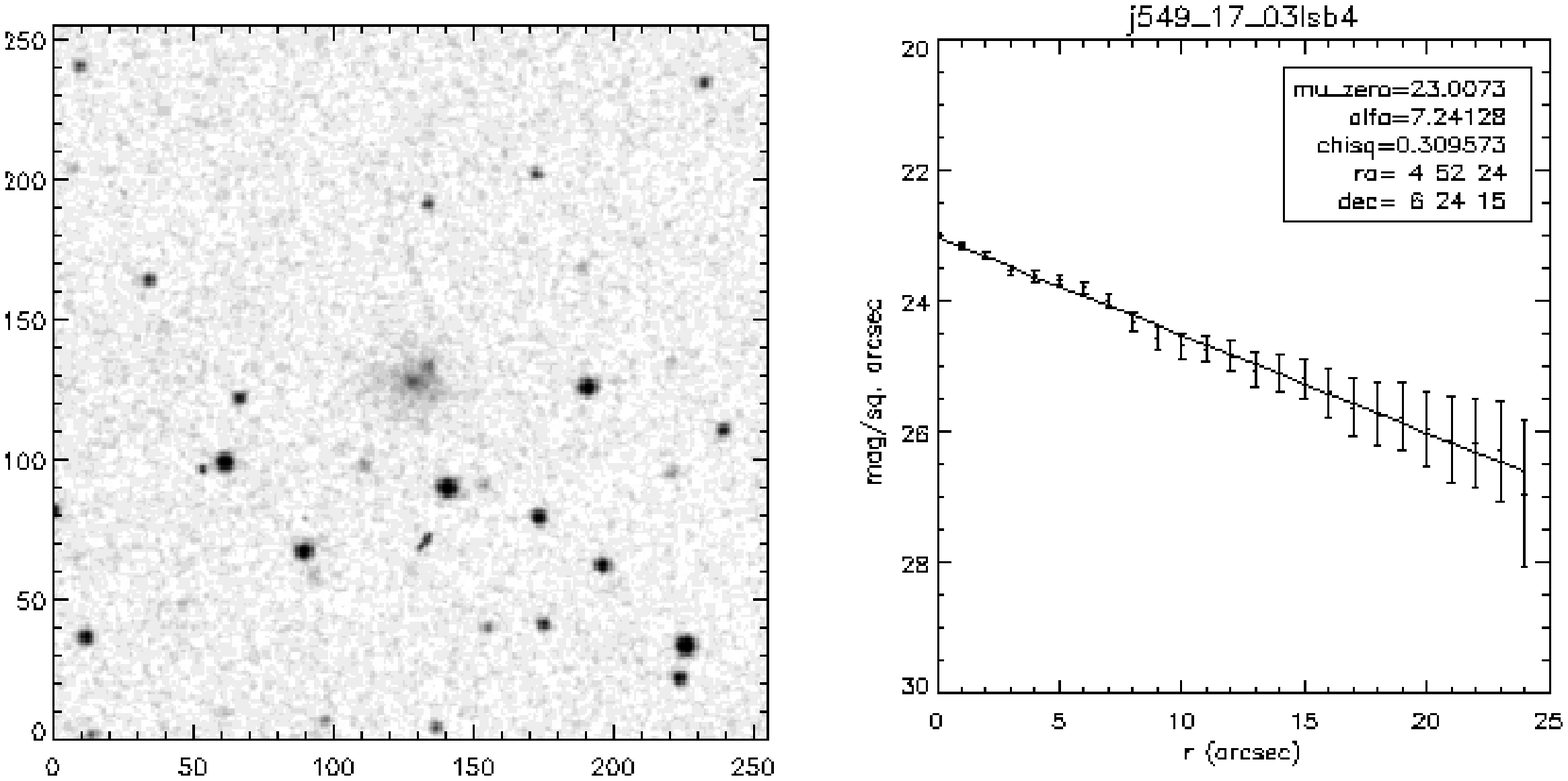} 
%for centering: act on hspace argument 
\hspace{9cm}
\epsfysize=6cm \epsfbox{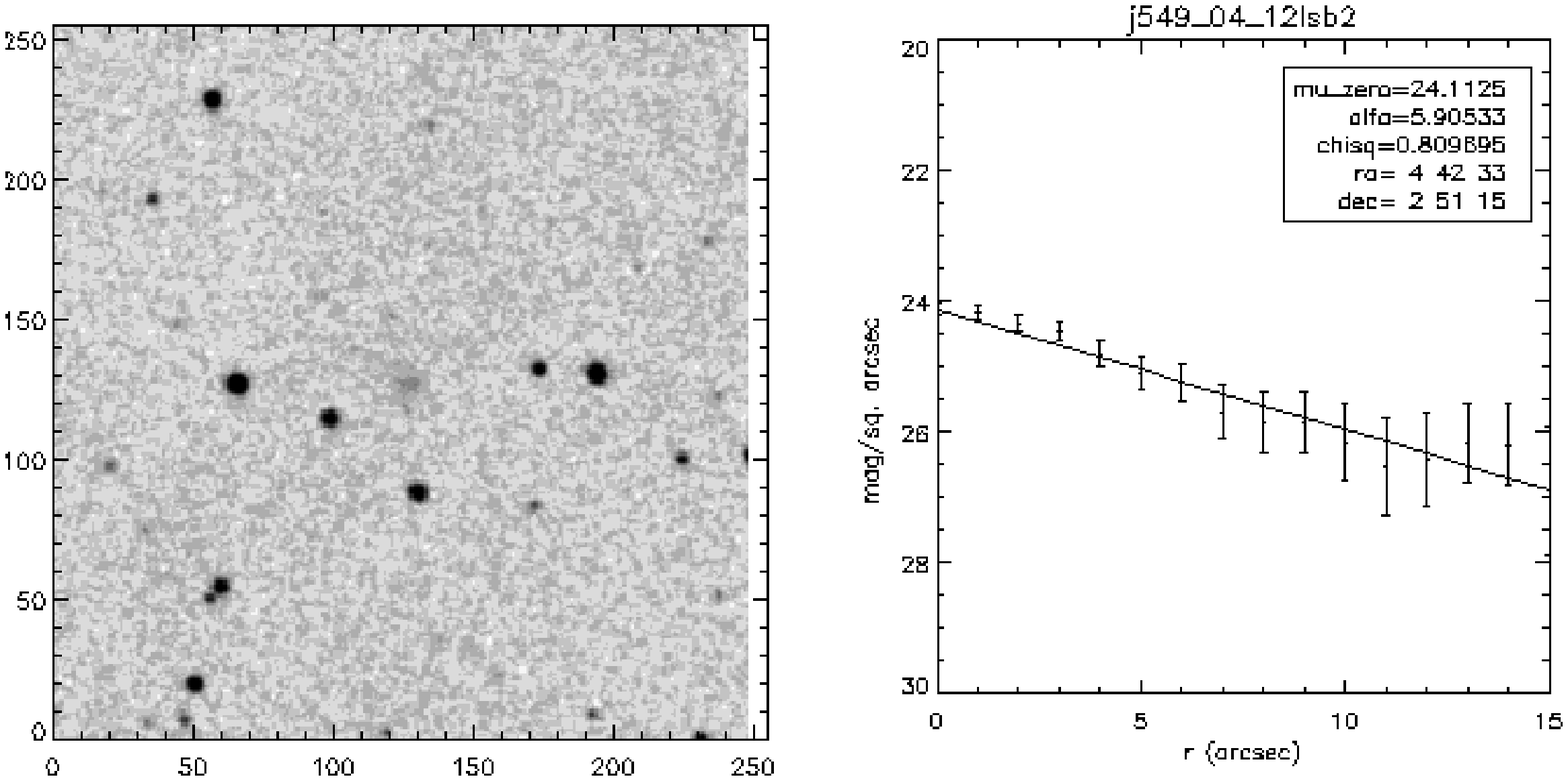}
\caption[h]{In this figure we show two of the LSBGs found by our
procedure on the field we have in common with Schombert $\&$ Bothun (1988), 
with their radial profiles.
The images and the profiles correspond to the J photographic band. 
The first galaxy is one of the objects already known (found in our 
significance map at 20 $\sigma$), while the second 
is one of the new objects, found at 5 $\sigma$.}{\label{fig3}}
\end{figure}
It is important to notice that one 
field we studied has already been analyzed in the first catalogue 
made by Schombert $\&$ Bothun (1988) and our algorithm finds the three known 
objects at high significance ($15 \div 20$ $\sigma$)
plus 6 other new ones (fig. \ref{fig3}).   
Our procedure can therefore detect even smaller and fainter objects than 
those until now catalogued on plates (see the solid line representing the 
detection limit of the procedure obtained by simulations, in fig \ref{fig2}, 
and the lined regions). 
Other advantages of the method are that (1) it is almost completely 
automatized and can easily manage wide field imaging data, as well as 
repeatable and objective detection; (2) the dataset (POSS II) covers the
 entire northern sky in 3 
filters (J,F,N) and this allow independent search in different filters 
(avoiding biases in colours) and the possibility to build a large catalog 
of LSBGs of known selection function, obtaining a significant statistical 
sample.

% References. We avoided using the \bibitem commmand since we found it is
% somewhat platform-dependent. We also avoided using the \cite{keyword}
% command since we found it cumbersome. However, if you are an expert 
% LateX user you may use the various LateX tools for the references 
% provided they give the same printout formats of the examples given here.

\end{document}